\documentclass[dvips]{siamltex}

\usepackage{amsfonts}
\usepackage{amsmath}
\usepackage{amssymb}
\usepackage{mathrsfs}
\usepackage{graphicx}
\usepackage[hypertex,colorlinks=true]{hyperref} % pour dvi
\usepackage{psfrag}

\newcommand{\HHH}{{\mathcal H }}

\newcommand{\NNN}{{\mathcal Z}}

\newcommand{\RR}{{\mathbb R}}

\newcommand{\EE}{{\mathbb E}}
\newcommand{\SSS}{{\mathbb  S}}

\newcommand{\tr}[1]{\text{Tr}\left(#1\right)}

\newcommand{\ket}[1]{\left|#1\right>}
\newcommand{\bra}[1]{\left<#1\right|}
\newcommand{\bket}[1]{\left<#1\right>}

\newtheorem{remark}{Remark}

\title{Real-time synchronization feedbacks for single-atom frequency standards \thanks{This work was partially supported
by the "Agence Nationale de la Recherche" (ANR), Projet Blanc CQUID
number 06-3-13957.}}

\author{
    Mazyar Mirrahimi\thanks{INRIA Rocquencourt, Domaine de
    Voluceau, Rocquencourt B.P. 105, 78153 Le Chesnay Cedex, France,
    ({\tt mazyar.mirrahimi@inria.fr}).}
    \and Pierre Rouchon \thanks{Mines ParisTech, Centre Automatique et Syst\`emes,
    60, bd. Saint-Michel, 75272 Paris Cedex 06, France, ({\tt
    pierre.rouchon@ensmp.fr}).} }

\begin{document}
\bibliographystyle{plain}
\maketitle

\begin{abstract}
Simple feedback loops, inspired from extremum-seeking, are proposed
to lock a probe-frequency to  the transition frequency of a single
quantum system following  quantum Monte-Carlo trajectories. Two
specific quantum systems are addressed, a 2-level one  and a 3-level
one that appears in coherence population trapping and optical
pumping. For both systems, the feedback algorithm is shown to be
convergent in the following sense: the probe frequency converges in
average towards the system-transition one and its standard deviation
can be made arbitrarily small. Closed-loop simulations  illustrate
robustness versus jump-detection efficiency and modeling errors.
\end{abstract}



\section{Introduction}\label{sec:intro}

\begin{figure}[htp]
  % Requires \usepackage{graphicx}
 \centerline{ \includegraphics[width=.5\textwidth]{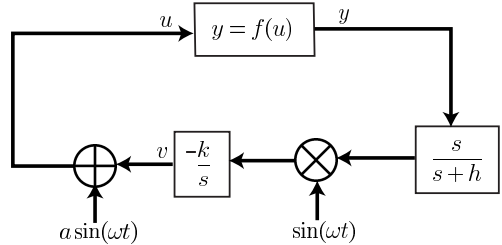} }
  \caption{The basic extremum seeking feedback loop for a non-linear static system $y=f(u)$ ($s=\frac{d}{dt}$ is the Laplace variable and $(h,k,a,\omega)$ are constant design parameters). }\label{fig:extremumseeking}
\end{figure}

The SI second is defined to be ``the duration of 9 192 631 770
periods of the radiation corresponding to the transition between the
two hyperfine levels of the ground state of the caesium 133
atom''~\cite{second}. A primary frequency standard is a device that
realizes this definition. Extremum seeking techniques (see, e.g,
\cite{ariyur-krstic:book03} for a recent exposure) are usually  used
in high precision spectroscopy to achieve frequency lock with an
atomic transition frequency~\cite{riis-et-al:04,peik-schneider-tam:06}. For  micro atomic-clocks
\cite{kitching-et-al:nist00}  synchronization is achieved when the
output signal of a photo-detector is maximum (or minimum). This
characterizes perfect resonance  between the probe laser frequency
with the atomic one. As sketched on
figure~\ref{fig:extremumseeking}, such  synchronization feedback
schemes are based on a modulation of the probe frequency, the input
$u$, with a sinusoidal variation  $a\sin(\omega t)$ of small
amplitude $a$  and fixed (low) frequency $\omega$,  on a high-pass
filtering (transfer function $\frac{s}{s+h}$) of the photo-detector signal
(the output $y$), on a multiplier and  finally an  integrator giving
the mean input value.

In such synchronization scheme the system corresponds to a
population of identical quantum systems with few  mutual
interactions (the vapor cell)  having reached its  asymptotic
statistical regime described by a density matrix solution of a
static Lindblad-Kossakovski  master equation. In this paper, we
propose to adapt this feedback strategy  to a single  quantum
system. Such a  system cannot be described by a  static non-linear
input/output map but it obeys a stochastic jump dynamics
\cite{dalibard-et-al:PRL92,haroche-raimond:book06}. The  output
signal is no more continuous since it corresponds  to  a counter
giving the jump times. As shown in~\cite{cohenT-dalibard:86},  all
the spectroscopic information and in particular the value of the
atomic transition frequency are  contained in the statistics of
these jump-time series. Thus  it is not surprising that such
feedback loops are possible. The contribution of this paper is to
propose for the first time (as far as we know)  a real-time
synchronization feedback scheme that can be implemented on
electronic circuits of similar complexity to those used for
extremum-seeking loops. In the  feedback loop, we avoid thus  the
use of quantum filters \cite{vanhandel:ieee05} and records of
jump-times sequences required by usual statistical treatments.

We consider  here two kinds of quantum systems. The first system is
the  simplest one we can imagine. It  has a stable ground state and
an excited unstable one. These two states are  in interaction with a
quasi-resonant electromagnetic field characterized  by a complex
amplitude $u+\imath v$ and a  frequency $\Omega$ close to the
transition frequency between the ground and excited states. The
measure corresponds then  to the  photon emitted by the excited
state when it relaxes to the ground state by spontaneous emission.
The complex amplitude is then  modulated according to $\bar u +
\imath \bar v \cos(\omega t)$ ($(\bar u,\bar v)$  positive
parameters, modulation frequency $\omega \ll \Omega$). The
synchronization  feedback  (playing the  role of the integrator in
figure~\ref{fig:extremumseeking}) corresponds essentially to  the
recurrence $\Omega_{N+1} = \Omega_N - \delta \sin(\omega t_N)$ where
$N$ is the jump-index, $t_N$ the jump-time, and $\Omega_N$ the probe
frequency between time $t_{N-1}$ and time $t_N$ ($\delta$ positive
parameter).  The second system corresponds to a typical
$\Lambda$-system appearing in  coherent population trapping
phenomena  and optical pumping~\cite{arimondo-96}. Such 3-level
configurations are also  present  in micro atomic clocks. The
synchronization feedback is  very similar to the previous one (see
subsection~\ref{ssec:algolambda}). Both feedback schemes are
illustrated by closed-loop  quantum Monte-Carlo simulations and rely
on two formal results (theorems~\ref{thm:qubit}
and~\ref{thm:lambda}) ensuring the convergence of the mean frequency
de-tuning  to $0$   with a standard deviation that can be made
arbitrary small.

The model considered in this paper corresponds to quantum Monte Carlo trajectories. Note that, although quantum trajectories represent a useful simulation method for the quantum master equation, they are interesting in their own right, since they model the measurement process itself and the resulting conditioned dynamics. In fact, some of the original work that motivated quantum trajectories~\cite{cohenT-dalibard:86,zoller-et-al-87} was to understand experiments on quantum jumps (for atoms with a vee configuration)~\cite{nagourney-et-al-86,sauter-et-al-86}. Therefore, through the paper, one should understand the ``quantum Monte Carlo'' trajectories as the model and not as a numerical method. We refer to~\cite{haroche-raimond:book06} for more details.

The paper is organized as follows. Section~\ref{sec:IntroQubit} is
devoted to the two-level system:  the stochastic jump dynamics are
depicted  in subsection~\ref{ssec:mathqubit}; the synchronization
feedback is detailed in subsection~\ref{ssec:algoqubit}; the
remaining two subsections deal with closed-loop simulations
illustrating theorem~\ref{thm:qubit}. Section~\ref{sec:IntroLambda}
deals with the $\Lambda$-system  and admits exactly the same
structure as section~\ref{sec:IntroQubit}. The two last
sections~\ref{sec:qubit} and~\ref{sec:lambda} are devoted to the
proofs of the two main results, theorems~\ref{thm:qubit}
and~\ref{thm:lambda}.

The authors thank Guilhem Dubois from LKB and Andr\'{e} Clairon from SYRTE for interesting discussions, suggestions and references in physics journals.

\section{The two-level system} \label{sec:IntroQubit}

   \subsection{Monte-Carlo trajectories}\label{ssec:mathqubit}

This 2-level system is defined on the Hilbert space $
\HHH=\text{span}\left\{\ket{g},\ket{e}\right\}$: the ground state
$\ket{g}$ is stable whereas the excited state $\ket{e}$ is unstable
with life time $1/\Gamma$ and relaxes to  $\ket{g}$.  The system is
submitted to a near-resonant laser field whose complex amplitude is
assumed to be slowly variable with respect to the transition
frequency. Its dynamics are stochastic  with quantum Monte-Carlo
trajectories \cite{haroche-raimond:book06} described  here below.

In the absence of quantum jump, the  density matrix $\rho$
evolves through the dynamics
$$
\frac{d}{dt}\rho=-\imath[\frac{H}{\hbar},\rho]-\frac{1}{2}\left\{L^\dag
L, \rho\right\} +\tr{L^\dag L \rho}\rho
$$
where $\left\{L^\dag L, \rho\right\}=L^\dag L \rho+ \rho L^\dag L$ stands for  the anti-commutator. Note that, the nonlinear term
in the above dynamics is being added to ensure that the density matrix remains normalized to trace 1. In fact, the usual formulation for
quantum Monte Carlo trajectories uses the wavefunction language and normalizes the wavefunction at each step. Here, without loss of generality and
for simplicity sakes we have translated these dynamics to the density matrix one.

The Hamiltonian
$\frac{H}{\hbar}=\frac{\Delta}{2}\sigma_z+u\sigma_x+v\sigma_y$ is
attached to the conservative part of the dynamics:
$(\sigma_x,\sigma_y,\sigma_z)$ are the Pauli matrices; $\Delta$
denotes the laser-atom detuning; $u$ and $v$ are the real
coefficients of the complex laser amplitude. The jump operator
$L=\sqrt{\Gamma}\ket{g}\bra{e}, $ is associated to the dissipative
dynamics with   $\Gamma>0 $ denoting  the decoherence rate. Note that,
the above Hamiltonian is the result of a common rotating wave approximation,
where the detuning and the decoherence rates are smaller than the transition
frequencies of the atom~\cite{haroche-raimond:book06}.

At each time step $dt$ the system may jump on the ground state
$\ket{g}\bra{g}$ with a probability given by
$$
p_{\text{jump}}(\rho\rightarrow \ket{g}\bra{g})=\tr{L^\dag
L\rho}dt=\Gamma\tr{\ket{e}\bra{e}\rho}dt =\Gamma~\bra{e}\rho\ket{e} dt
.
$$
Each jump is associated to the spontaneous  emission of a photon
that is detected  by the photo-detector: the measurement is just a
simple click and we know that just after the click the system is at
the ground state, i.e., $\rho=\ket g \bra g$.

In the sequel, we will use this stochastic dynamics in the $\Gamma$-scale. This just consists in  replacing
$u$ by $ u \Gamma$, $v$ by $v \Gamma$, $\Delta$ by $\Delta \Gamma$, and $t$ by $t /\Gamma$ in the equations. In this de-coherence time-scale,  the  density matrix $\rho$
evolves through the dynamics
\begin{equation}\label{eq:qubit}
\frac{d}{d\bar t}\rho=-\imath\left[\frac{\Delta}{2}\sigma_z+u\sigma_x+v\sigma_y,\rho \right]-\frac{1}{2}\left\{\ket e \bra e, \rho\right\} + \bra e \rho \ket e \rho
\end{equation}
and the  jump probability between $t$ and $t+dt$ reads
\begin{equation}\label{eq:jumpqubit}
p_{\text{jump}}(\rho\rightarrow \ket{g}\bra{g})=\bra{e}\rho\ket{e} dt
.
\end{equation}
Just after each jump, $\rho$ coincides with $\ket{g}\bra{g}$. The
atom/laser detuning is $\Delta$ and the laser complex amplitude is
$u+\imath v$.

\subsection{The synchronization feedback}\label{ssec:algoqubit}

We consider  here the two-level system described, in the decoherence
time-scale,  by~\eqref{eq:qubit}~\eqref{eq:jumpqubit}. The quantum
jumps lead to the emission of photons that will be detected with a
certain efficiency $\eta\in (0,1]$.

The main goal of this paper is to provide a real-time algorithm so
that, using the information obtained through the detected photons,
we can synchronize the laser with the atomic transition frequency and therefore
make $\Delta$ converge to zero.

Note that, in practice we have a certain knowledge of the transition
frequency and therefore, we can always tune our laser so that the
detuning $|\Delta|$ does not get larger than a fixed constant $C$.

In the aim of providing a synchronization algorithm inspired from extremum-seeking, we consider a laser field  amplitude of the form
$$
u=\bar u , \quad  v=
\bar v\cos(\omega t)
$$
where the modulation frequency $\omega$ is of order $1$ but where $\bar u $ and $\bar v$ are small:
$\bar  u, \bar  v\ll 1$.

The main strategy for the correction of the detuning is to wait for
the matured quantum jumps (clicks of the photo-detector). This means
that we choose a certain time constant $T \gg 1$ and if the distance
between two jumps is more than $T$, we will correct the detuning
according to the time when the second jump happens. Note that, one
can easily show that these matured quantum jumps, almost surely,
happen within a finite horizon. Here is the explicit feedback
algorithm:
\begin{enumerate}
  \item Start with a certain detuning $\Delta_0$ with
  $|\Delta_0| \leq C$ and set the switching parameter $S=0$ and the counter $N=0$.
  \item Wait for a first click and meanwhile evolve the switching parameter through
  $\frac{d}{dt}S=1$.
  \item If the click
  happens while $S\leq T$ then switch the parameter $S$ to zero and
  go back to the step 2.
  \item If the click
  happens while $S> T$ then switch the parameter $S$ to zero, change the counter value to
  $N+1$,
  correct the detuning $\Delta_N$ as follows:
  $$
  \left\{
  \begin{aligned}&
  \Delta_{N+1}=\Delta_N-\delta\sin(\omega t)\qquad
  &\text{if } |\Delta_N-\delta\sin(\omega t)|\leq C,\\
  &\Delta_{N+1}=C,\qquad &\text{otherwise}
  \end{aligned}
  \right.
  $$
  and go back to the step 2.
\end{enumerate}
Here we have chosen the correction gain $\delta \ll 1$. Our claim is
that such an algorithm provides an approximate synchronization of laser frequency:  given any small $\epsilon$, we can
adjust the design parameters $\bar u$, $\bar v$ and $\bar \delta$
small enough such that with the above algorithm, the detuning
$\Delta_N$ converges in average to an $O(\epsilon^2)$-neighborhood
of 0 with a deviation of order $O(\epsilon)$ (in the
$\Gamma$-scale): according to theorem~\ref{thm:qubit}, it suffices
to take $\bar u, \bar v \sim \epsilon$, $\delta \sim \epsilon^2$ and
$(\omega,C)$ such that $ 4 \omega^2 > 1 + 4 C^2$ to ensure such
convergence. In theorem~\ref{thm:qubit}, the "dead-time" $T$ between two jumps is chosen  mostly for
technical reasons during the proof of theorem~\ref{thm:qubit}. It is
related to the  convergence time for the jump-free
dynamics~\eqref{eq:qubit}  starting with $\ket g \bra g $  towards
an $\epsilon^4$-neighborhood of its asymptotic regime. Since the
convergence is exponential, $T$ is linear in $-\log\epsilon$, this
explain the fact that we can choose $T$ around $1$, even if
$\epsilon$ is very small. However, in simulation, we have observed no
convergence difference between $T>0$ large (around $10$) and $T=0$.
Therefore, this dead time does  not seem to be necessary
in practice and one can take it simply to be 0.
Finally, notice that, such algorithm is very simple and can be
implemented via a  standard electronic circuit. Indeed, the feedback loop
only adapts  the atom/laser  detuning $\Delta$ that is of several orders of
magnitude smaller that the optical transition  frequency involving femto-second time-scales.
For the two-level system, the adaptation update frequency is  directly related  to the time-interval
between detector clicks and  thus is less than  $\Gamma$, the inverse of the life time of the unstable
excited state. Since, in several physical situations \cite{peik-schneider-tam:06,riis-et-al:04}, $\Gamma$
lies in the radio frequency domain,  the adaptation algorithm can be realized via a simple analogue circuit in the GHz range.

\subsection{Numerical simulations}

\begin{figure}[h]\psfrag{T}{\footnotesize{Time}}\psfrag{N}{\footnotesize{Number of clicks}}\psfrag{L}{\footnotesize{Laser detuning}}
\begin{center}
\centerline{\includegraphics[width=1.2\textwidth]{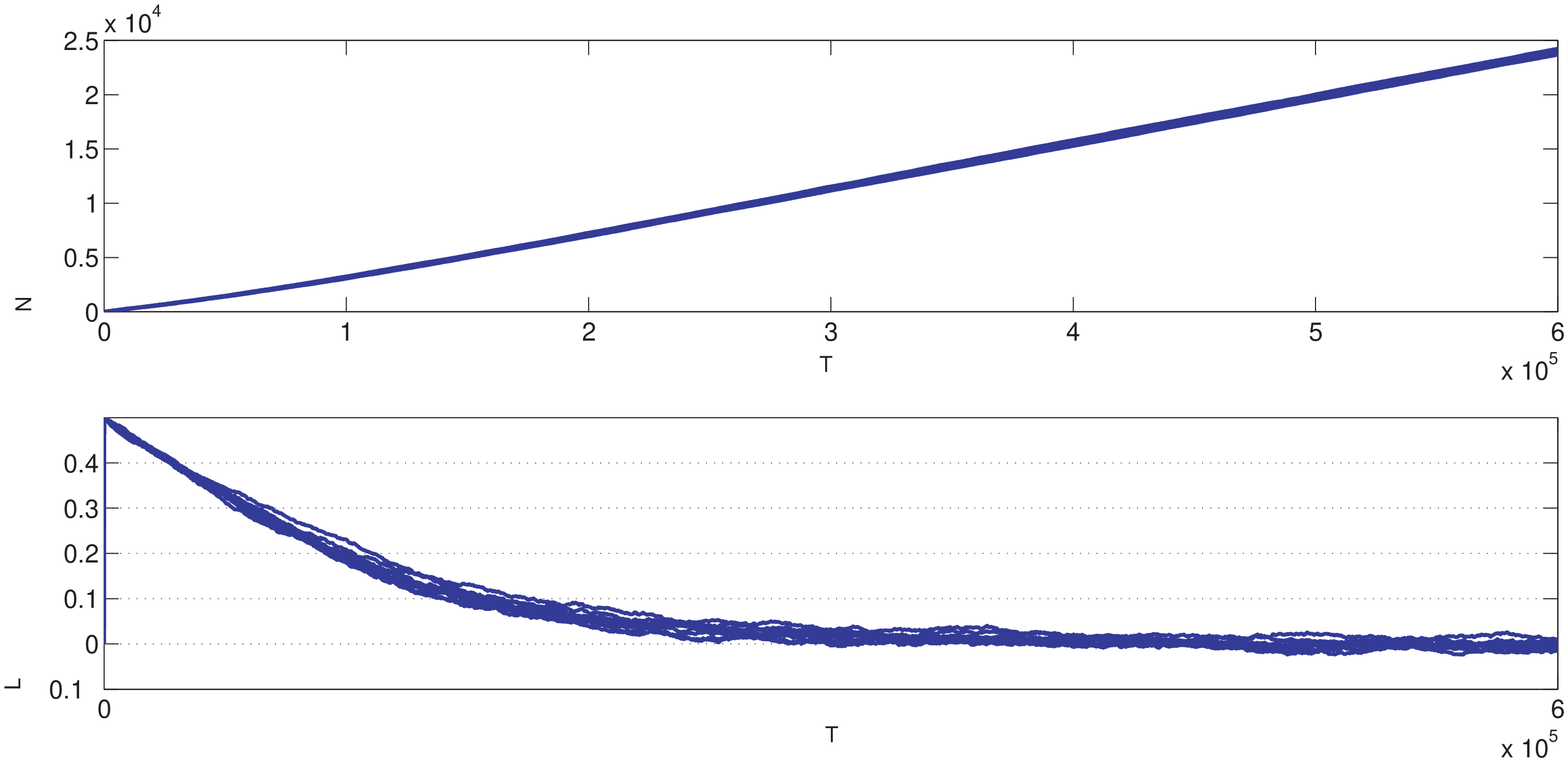}}
\caption{The detuning evolution versus the number of quantum jumps
for the synchronization algorithm of
Subsection~\eqref{ssec:algoqubit}.}\label{fig:qubit}
\end{center}
\end{figure}

Let us now show the performance of this algorithm on some
simulations. For the simulations of Figure~\ref{fig:qubit},  we take ($\Gamma=1$ in the decoherence time-scale)
$$
C=1/2,\quad  \eta=0.9, \quad  \bar u=\bar v = 6.0~10^{-2}, \quad
\omega=1.0 , \quad \delta=9.0~10^{-4}. $$ Figure~\ref{fig:qubit}
correspond to  10 random trajectories of the system starting with
the same initial condition  $\rho_0=\ket{g}\bra{g}$ and detuning
$\Delta_0=C=1/2$.
 The first plot provides the number of clicks (quantum jumps) while the second one gives the
evolution of the detuning $\Delta_N$. As it can be noted, the
detuning converge to $0$ in average with a standard deviation of
order $\epsilon$ (here  $\epsilon\sim  10^{-2}$). In these
simulations,  we take the parameters $T=0$.

\subsection{Formal result} The proof of the following theorem underlying  the above simulations is given in section~\ref{sec:qubit}.
 we have the following theorem
\begin{theorem}\label{thm:qubit}
Consider the Monte-Carlo trajectories  described  by~\eqref{eq:qubit}-\eqref{eq:jumpqubit}. Assume  a perfect  jump-detection efficiency $\eta=1$ and take  the synchronization-feedback  presented in
subsection~\ref{ssec:algoqubit}  with
\begin{equation}\label{eq:eps1}
\overline u, \overline v\sim \epsilon\ll 1.
\end{equation}
Assume that
\begin{equation}\label{eq:eps2:C}
\delta\sim \epsilon^2,\quad
4C^2+1<4\omega^2.
\end{equation}
We can fix then the dead-time  $T$ in the algorithm large enough
so that:
\begin{equation}\label{eq:result1}
\limsup_{N\rightarrow\infty} \EE(\Delta_N^2)\leq
O(\epsilon^2).
\end{equation}
\end{theorem}

\begin{corollary}\label{cor:qubit}
Under the assumptions of the Theorem~\ref{thm:qubit}, one has
\begin{equation}\label{eq:res1}
\limsup_{N\rightarrow \infty} P(|\Delta_N|>\sqrt{\epsilon})\leq
O(\epsilon).
\end{equation}
\end{corollary}
This corollary results from
the Markov inequality:
$$
P(|\Delta_N|>\sqrt{\epsilon})= P(\Delta_N^2>\epsilon)\leq
\frac{\EE(\Delta_N^2)}{\epsilon}.
$$
Therefore applying~\eqref{eq:result1}, one deduces~\eqref{eq:res1}.

%\begin{remark}\label{rem:exp}
%Following the steps of the proof in Section~\ref{sec:qubit}, one may
%also show that:
%\begin{equation*}
%\limsup_{N\rightarrow\infty} |\EE(\Delta_N)|\leq O(\epsilon^2).
%\end{equation*}
%\end{remark}

\begin{remark}\label{rem:C}
The assumption~\eqref{eq:eps2:C} is not so restrictive. Indeed, for
an a priori knowledge of the detuning magnitude, by taking a large
enough frequency $\omega$, one can ensure the relevance of this
assumption.
\end{remark}

\section{The $\Lambda$-system}\label{sec:IntroLambda}

   \subsection{Monte-Carlo trajectories} \label{ssec:mathlambda}

Here, we consider a three time-scale system where a laser irradiates
a 3-level $\Lambda$-system. The system is composed of 2 (fine or
hyperfine) ground states $\ket{g_1}$ and $\ket{g_2}$ having energy
separation in the radio-frequency or microwave region, and an
excited state $\ket{e}$ coupled to the lower ones by optical
transitions at frequencies $\omega_1$ and $\omega_2$. The
decay times for the optical coherences are assumed to be much
shorter than those corresponding to the ground state transitions
(here assumed to be metastable).

\begin{figure}[h]
\begin{center}
\centerline{\includegraphics[width=.4\textwidth]{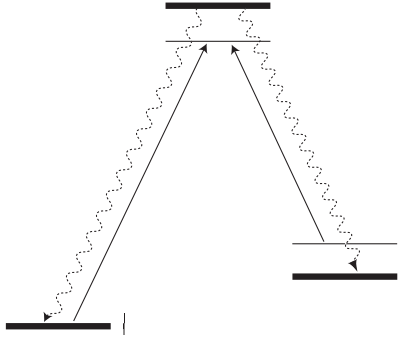}}
\caption{Relevant energy levels, transition and decoherence terms
for the $\Lambda$-system~\eqref{eq:lam}.}\label{fig:LambdaSyst}
\end{center}
\end{figure}

Applying near-resonant laser fields and under the rotating wave
approximations, while assuming the transition frequencies $\omega_1$
and $\omega_2$ much higher than the other frequencies, we can remove
one of the time scales. Then, the quantum Markovian master equation
of Lindblad type, modeling the evolution of a statistical ensemble
of identical systems given by Figure~\ref{fig:LambdaSyst}, reads
(see~\cite{haroche-raimond:book06}, chapter 4, for a tutorial and
exposure on such master equation):
\begin{equation}\label{eq:lam}
\frac{d}{dt}\rho=
-\frac{\imath}{\hbar}[\tilde H,\rho]+\frac{1}{2}\sum_{j=1}^2(2Q_j\rho
Q_j^\dag-Q_j^\dag Q_j\rho-\rho Q_j^\dag Q_j),
\end{equation}
where
\begin{multline*}
\frac{\tilde H}{\hbar}=\frac{\Delta}{2}(\ket{g_2}\bra{g_2}-\ket{g_1}\bra{g_1})
+\left(\Delta_e+\frac{\Delta}{2}\right)(\ket{g_1}\bra{g_1}+\ket{g_2}\bra{g_2})\\
+\widetilde\Omega_1\ket{g_1}\bra{e}+\widetilde\Omega_1^\ast\ket{e}\bra{g_1})
+\widetilde\Omega_2\ket{g_2}\bra{e}+\widetilde\Omega_2^\ast\ket{e}\bra{g_2})
\end{multline*}
and $ Q_j=\sqrt{\Gamma_j}\ket{g_j}\bra{e}$. Here, $\Delta$
represents the Raman detuning, $|\widetilde\Omega_1|$ and
$|\widetilde\Omega_2|$ are the so-called Rabi frequencies and
$\Gamma_1$ and $\Gamma_2$ are decoherence rates.
Assuming the  decoherence rates $\Gamma_1$ and $\Gamma_2$ much
larger than  the Rabi frequencies $|\widetilde\Omega_1|$,
$|\widetilde\Omega_2|$, and the detuning frequencies  $\Delta$ and
$\Delta_e$, we may apply the singular perturbation theory to remove
these fast and stable dynamics. Indeed, the dynamics corresponding
to the excited state $\ket{e}$ represent the fast dynamics and can
be removed in order to obtain a system living on the 2-level
subspace $\text{span}\{\ket{g_1},\ket{g_2}\}$. The reduced Markovian
master equation is still of Lindblad type and reads
(see~\cite{mirrahimi-rouchon:ieee08} for a detailed proof)
\begin{equation}\label{eq:lam2}
\frac{d}{dt}\rho=
-\frac{\imath}{\hbar}[H,\rho]+\frac{1}{2}\sum_{j=1}^2(2 L_j\rho
L_j^\dag- L_j^\dag  L_j\rho-\rho  L_j^\dag  L_j),
\end{equation}
where the reduced slow-Hamiltonian $H$ is given, up-to a global phase change, by
\begin{equation}\label{eq:lamaux}
\frac{H}{\hbar}=\frac{\Delta}{2}(\ket{g_2}\bra{g_2}-\ket{g_1}\bra{g_1})
=\frac{\Delta}{2} \sigma_z
\end{equation}
and
\begin{equation}\label{eq:lam2aux}
L_j=\sqrt{\tilde \gamma_j}\ket{g_j}\bra{b_{\widetilde\Omega}}\qquad\text{with}\qquad
\tilde \gamma_j=4\frac{|\widetilde\Omega_1|^2+|\widetilde\Omega_2|^2}{(\Gamma_1+\Gamma_2)^2}\Gamma_j.
\end{equation}
Here, $\ket{b_{\tilde \Omega}} $ represents the bright state (in the coherent
population trapping)
$$
\ket{b_{\widetilde\Omega}}=\frac{\widetilde\Omega_1}{\sqrt{|\widetilde\Omega_1|^2+|\widetilde\Omega_2|^2}}\ket{g_1}
+\frac{\widetilde\Omega_2}{\sqrt{|\widetilde\Omega_1|^2+|\widetilde\Omega_2|^2}}\ket{g_2}.
$$
From now on, we deal with the 2-level system~\eqref{eq:lam2} instead
of~\eqref{eq:lam}.

In order to characterize the Monte-Carlo trajectories of the system,
we note that in the absence of the quantum jumps the reduced slow
system evolves through the dynamics:
$$
\frac{d}{dt}\rho=-\imath\frac{\Delta}{2}[\sigma_z,\rho]-\frac{1}{2}\sum_{j=1}^2\left\{L_j^\dag
L_j, \rho\right\} +\sum_{j=1}^2\tr{L_j^\dag L_j \rho}\rho,
$$
the Lindblad operators $L_j$ being given by~\eqref{eq:lam2aux}.
Since $L_j^\dag L_j= \tilde \gamma_j
\ket{b_{\widetilde\Omega}}\bra{b_{\widetilde\Omega}}$ we have, with
$\tilde \gamma = \tilde \gamma_1+\tilde \gamma_2$,
\begin{equation}\label{eq:lambdaOld}
   \frac{1}{\tilde\gamma} \frac{d}{dt}\rho=
   -\imath\frac{\Delta}{2\tilde\gamma}[\sigma_z,\rho]
   -\frac{1}{2}\left\{ \ket{b_{\widetilde\Omega}}\bra{b_{\widetilde\Omega}}, \rho\right\} +\tr{\ket{b_{\widetilde\Omega}}\bra{b_{\widetilde\Omega}} \rho}\rho.
\end{equation}
At each time step $dt$ the system may jump towards the state
$\ket{g_j}\bra{g_j}$ with a probability given by:
\begin{equation}\label{eq:propjump}
P_{\text{jump}}(\rho\rightarrow \ket{g_j}\bra{g_j})=\tr{L^\dag_j
L_j\rho}dt=\tilde\gamma_j\tr{\ket{b_{\widetilde\Omega}}\bra{b_{\widetilde\Omega}}}dt,\quad
j=1,2.
\end{equation}
As it can be seen this probability is proportional to the population
of the bright state $\ket{b_{\widetilde\Omega}}$ (this is actually
the reason to call $\ket{b_{\widetilde\Omega}}$ the bright state).

   \subsection{The synchronization feedback}\label{ssec:algolambda}

In this subsection, we consider the 2-level system as the slow
subsystems of the $\Lambda$-system presented in
subsection~\ref{ssec:mathlambda}. The only change we admit is that
instead of constant amplitude laser fields $\widetilde\Omega_1$ and
$\widetilde\Omega_2$, we consider amplitudes varying with a
frequency much lower than the decoherence rate $\Gamma_1$ and
$\Gamma_2$. Consider two positive constant Rabi-frequencies
$\Omega_1$ and $\Omega_2$
 ($\Omega_1, \Omega_2 \ll \Gamma_1, \Gamma_2$) and take   the following  modulations
\begin{equation}\label{W1W2mod:eq}
    \widetilde\Omega_1 = \Omega_1 + \imath \epsilon \Omega_2\cos(\omega t), \quad
    \widetilde\Omega_2 = \Omega_2 - \imath \epsilon \Omega_1 \cos(\omega t)
\end{equation}
with $\epsilon \ll 1$ and $\omega \ll \Gamma_1, \Gamma_2$.
Following subsection~\ref{ssec:mathlambda}, consider the orthogonal  basis
\begin{equation}\label{eq:bd}
  \ket{b} = \frac{\Omega_1 \ket{g_1} + \Omega_2 \ket{g_2}}
                 {\sqrt{\Omega_1^2+\Omega_2^2}}
  ,\quad
  \ket{d} = \frac{\Omega_2 \ket{g_1} - \Omega_1 \ket{g_2}}
                 {\sqrt{\Omega_1^2+\Omega_2^2}}
\end{equation}
and set
\begin{equation}\label{eq:g1g2}
    \gamma_j =4\frac{\Omega_1^2+ \Omega_2^2}{(\Gamma_1+\Gamma_2)^2}\Gamma_j, \quad \text{for~}j=1,2 \quad \text{and~} \gamma=\gamma_1 + \gamma_2
    .
\end{equation}
Here, $\ket{b}=\ket{b_{\Omega}}$  (resp. $\ket{d}=\ket{d_{\Omega}}$)
denote  the bright (resp.  dark) state of the unperturbed
non-oscillating system.

If we replace $\Delta/\gamma$ by $\Delta$, $\omega/\gamma$ by
$\omega$ and $\gamma t$ by $t$ in the stochastic
dynamics~\eqref{eq:lambdaOld} and jump
probability~\eqref{eq:propjump}, we get the quantum jump dynamics in
the $1/\gamma$ scale, the optical-pumping scale, that reads:
\begin{itemize}
\item In the absence of quantum jumps, the systems density matrix $\rho$
evolves through the dynamics
\begin{multline} \label{eq:lambda}
\frac{d}{dt}\rho=-\imath\left[\frac{\Delta}{2}\sigma_z,\rho\right]
-\frac{1}{2}\left\{\ket{b+\iota\epsilon\cos(\omega t)
d}\bra{b+\iota\epsilon\cos(\omega t)
d},\rho\right\}\\+\tr{\ket{b+\iota\epsilon\cos(\omega t)
d}\bra{b+\iota\epsilon\cos(\omega t) d}\rho}\rho.
\end{multline}
with $\ket b =\cos\alpha \ket{g_1} + \sin\alpha \ket{g_2}$, $\ket d
=-\sin\alpha \ket{g_1} + \cos\alpha \ket{g_2}$
($\alpha\in\left[0,\frac{\pi}{2}\right]$ is the argument of
$\Omega_1 + \imath \Omega_2$).

\item
At each time step $dt$ the system may jump on the ground state
$\ket{g_j}$ ($j=1,2$) with a probability given by
\begin{equation}\label{eq:jump}
p_{\text{jump}}(\rho\rightarrow \ket{g_j}\bra{g_j})=
\frac{\gamma_j}{\gamma}\tr{\ket{b+\imath\epsilon \cos(\omega
t)d}\bra{b+\imath\epsilon \cos(\omega t)d}\rho}dt
\end{equation}
This quantum jump leads to the emission of a photon that will be
detected with certain efficiencies: $\eta_j\in (0,1]$ for the jumps
to the state $\ket{g_j}$.

\end{itemize}
We assume a broad band detection process and thus the only information available with such measure  is just the jump time. The type of jump (either to $\ket{g_1}$ or $\ket{g_2}$) is not available here.
Thus the total jump probability reads
\begin{equation}\label{eq:jumptot}
p_{\text{jump}}=\tr{\ket{b+\imath\epsilon \cos(\omega
t)d}\bra{b+\imath\epsilon \cos(\omega t)d}\rho}dt
\end{equation}
After each jump, $\rho$ coincides with $\ket{g_1}\bra{g_1}$ or $\ket{g_2}\bra{g_2}$.

Similarly to the last subsection, we are interested in synchronizing
the lasers with the system's frequencies and therefore make $\Delta$
converge to zero. As for the  two-level case, we have a certain
knowledge of the system's frequencies and therefore, we can always
tune our lasers so that the detuning $|\Delta|$ does not get larger
than a fixed constant $C$.

Assume that $
\epsilon\ll 1 \ll \omega$ and consider the following
synchronization algorithm:
\begin{enumerate}
  \item Start with a certain detuning $\Delta_0$ with
  $|\Delta_0| \leq C$ and set the switching parameter $S=0$ and the counter $N=0$.
  \item Wait for a first click and meanwhile evolve the switching parameter through
  $\frac{d}{dt}S=1$.
  \item If the click
  happens while $S\leq T$ then switch the parameter $S$ to zero and
  go back to the step 2.
  \item If the click
  happens while $S> T$ then switch the parameter $S$ to zero, change the counter value to
  $N+1$,
  correct the detuning $\Delta_N$ as follows:
  $$
  \left\{
  \begin{aligned}&
  \Delta_{N+1}=\Delta_N-\delta \sin(2\alpha)\cos(\omega t)\qquad
  &\text{if } |\Delta_N-\delta \sin(2\alpha)\cos(\omega t)|\leq C,\\
  &\Delta_{N+1}=C,\qquad &\text{otherwise}
  \end{aligned}
  \right.
  $$
  and go back to the step 2.
\end{enumerate}
Here, we have chosen the correction gain $\delta\ll 1$. Similarly to
the last subsection, we claim that, given any small $\epsilon$, we
can adjust the parameters $\omega$ large and $\delta$ small enough
such that with the above algorithm, the detuning $\Delta_N$
converges in average to an $O(\epsilon^2)$-neighborhood of 0 with a
deviation of order $O(\epsilon)$. Here again, the time constant $T$ is a technical parameter and is
necessary for the proof of the theorem. The numerical simulations illustrate that this is not necessary
in practice and one can take it simply to be 0.

   \subsection{Numerical simulations}

\begin{figure}[h]\psfrag{T}{\footnotesize{Time}}\psfrag{N}{\footnotesize{Number of clicks}}\psfrag{R}{\footnotesize{Raman detuning}}
\begin{center}
\centerline{\includegraphics[width=1.2\textwidth]{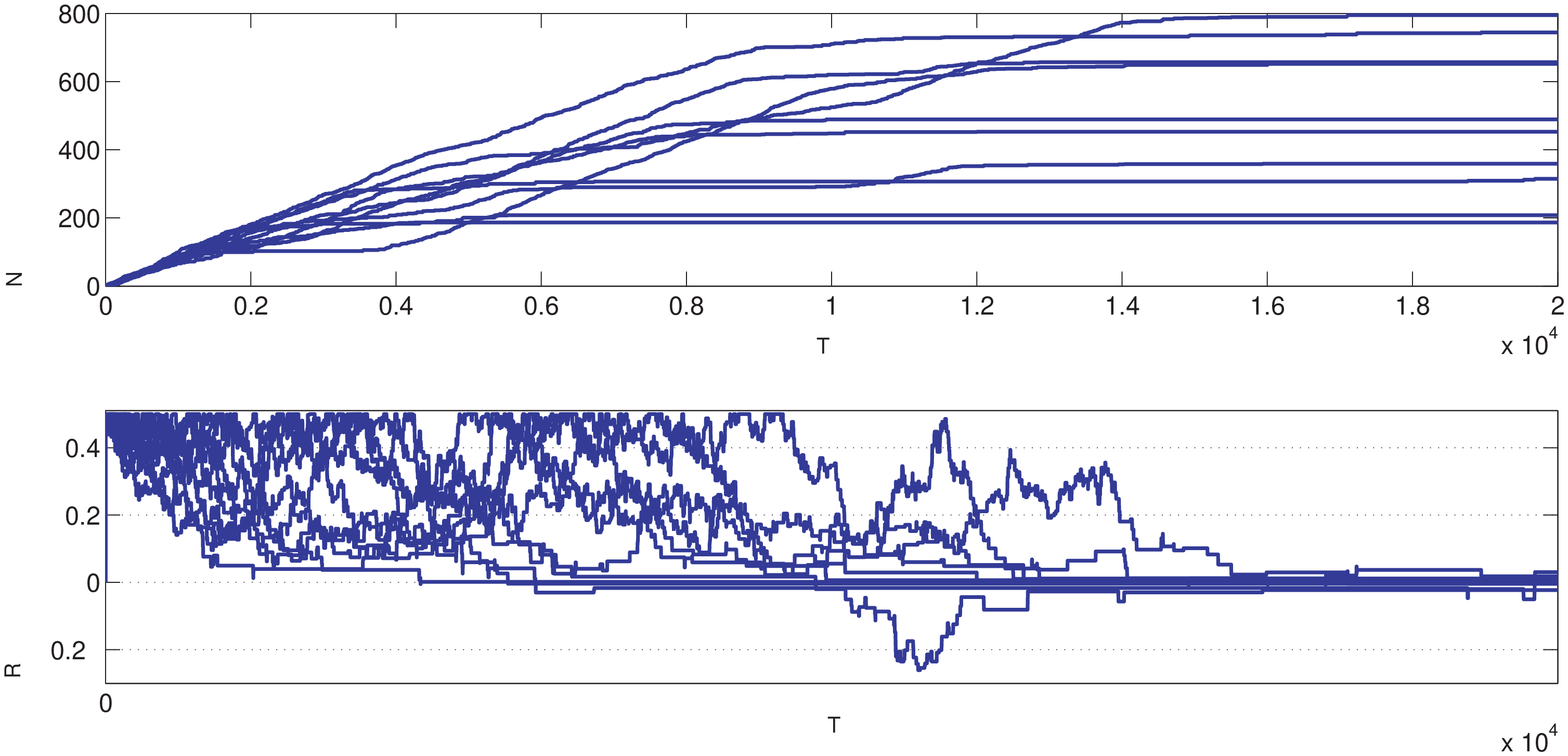}}
\caption{The detuning evolution versus the number of quantum jumps
for the synchronization algorithm of
Subsection~\eqref{ssec:algoqubit}.}\label{fig:lambda}
\end{center}
\end{figure}

Let us now show the performance of this algorithm on some
simulations. In the simulations of Figure~\ref{fig:lambda}, we apply
the above synchronization strategy directly on the main
$\Lambda$-system (and not on the slow 2-level subsystem).

We take the parameters $C=0.5$, $\Omega_1=\Omega_2=1$ (i.e.
$\alpha=\pi/4$), $\Gamma_1=\Gamma_2=3.0$ (i.e.
$\gamma_1=\gamma_2=0.6667$), $\eta_1=0.9$, $\eta_2=1.0$,
$\epsilon=0.03$, $\gamma/\omega=0.05$ and $\delta=0.015$. The
simulations of Figure~\ref{fig:qubit}, then, illustrate 10 random
trajectories of the system starting at $\Delta_0=.5$ and
$\rho_0=\ket{d}\bra{d}$ where $\ket{d}=\frac{1}{\sqrt
2}(\ket{g_1}-\ket{g_2})$. The first plot provides the number of
clicks (quantum jumps) while the second one gives the evolution of
the detuning $\Delta_N$. As it can be noted, the detuning converges
to a small neighborhood of zero within at most $1000$ clicks.

\subsection{Formal result} The proof of the following theorem is given in section~\ref{sec:lambda}.

\begin{theorem}\label{thm:lambda}
Consider the Monte-Carlo trajectories described by~\eqref{eq:lambda}-\eqref{eq:jump}  where
\begin{equation}\label{eq:condb}
\ket{b}=\cos\alpha\ket{g_1}+\sin \alpha\ket{g_2}~\quad \text{with} \quad 0 < \alpha < \frac{\pi}{2}.
\end{equation}
Moreover, we assume perfect detection efficiency  $\eta_1=\eta_2=1$ and
\begin{equation}\label{eq:lameps1}
\epsilon\ll 1,\qquad \frac{1}{\omega}\sim\epsilon^{2}.
\end{equation}
Consider then the synchronization algorithm of subsection~\ref{ssec:algolambda}
with
\begin{equation}\label{eq:lameps2}
C < 1/2 \qquad\text{and}\qquad\delta\sim \epsilon^3.
\end{equation}
We can fix then the time constant $T$ in the algorithm large enough
so that:
\begin{equation}\label{eq:result3}
\limsup_{N\rightarrow\infty} \EE(\Delta_N^2)\leq O(\epsilon^2).
\end{equation}
\end{theorem}

\begin{remark}\label{rem:eps}
Following the steps of the proof and changing the assumptions
~\eqref{eq:lameps1} and~\eqref{eq:lameps2} to
\begin{equation}\label{eq:lameps3}
1/\omega\sim\epsilon\qquad\text{and}\qquad\delta\sim\epsilon^2,
\end{equation}
one can show that, the detuning reaches an
$O(\epsilon)$-neighborhood of $0$ with a deviation of order
$\sqrt{\epsilon}$.

This is actually the assumption~\eqref{eq:lameps3} that is relevant
for the real system and that is considered in the simulations of the
subsection~\ref{ssec:algolambda}. In fact, through this assumption
the slow/fast approximation of~\cite{mirrahimi-rouchon:ieee08} is
still available and therefore the system~\eqref{eq:lambda} is a
relevant approximation of the real $\Lambda$-system.
\end{remark}

\begin{remark}\label{rem:lambda}
Similarly to the two-level case, we have
\begin{equation*}
\limsup_{N\rightarrow \infty} P(|\Delta_N|>\sqrt{\epsilon})\leq
O(\epsilon).
\end{equation*}
\end{remark}

\section{Proof of theorem~\ref{thm:qubit}}\label{sec:qubit}
 In order to simplify the notations, we
assume
$$
\overline u=\epsilon \kappa_1, \qquad \overline v=\epsilon \kappa_2,
\qquad \delta=\epsilon^2\kappa_3,
$$
where $\kappa_1,\kappa_2,\kappa_3\sim 1$.

We proceed the proof of the Theorem~\ref{thm:qubit} in two main
steps:
\begin{description}
  \item[Step 1] We consider the evolution in the absence of the
  quantum jumps through the system~\ref{eq:qubit}. We study the
  asymptotic regime of the dynamics. The constant time $T$ will then be chosen to
  ensure the non-jumping system to reach an $\epsilon^4$-neighborhood
  of the limit regime.
  \item[Step 2] In the second step, applying the result of the first step,
  we calculate the conditional expectation of $\Delta_{N+1}$ having fixed $\Delta_N$.
  Finally, we sum up all these results in order to find the limit~\eqref{eq:result1}.
\end{description}

\subsection{Step 1: asymptotic regime of the non-jumping
system}\label{ssec:step1qubit}

We are interested in the dynamics of the system
\begin{multline}\label{eq:main}
\frac{d}{dt}\rho=-\imath\left[\frac{\Delta}{2}\sigma_z,\rho\right]
-\frac{1}{2}\left\{\ket{e}\bra{e},\rho\right\}+\tr{\ket{e}\bra{e}\rho}\rho\\
+\epsilon\left[\kappa_1 \sigma_x+\kappa_2\cos(\omega
t)\sigma_y,\rho\right].
\end{multline}
In this aim, we apply the averaging theorem~(see
e.g.~\cite{guckenheimer-holmes-book}, page 168). The un-perturbed
dynamics, given by the first line of~\eqref{eq:main}, admits an
asymptotically stable hyperbolic equilibrium given by
$\ket{g}\bra{g}$. Therefore, applying the averaging theorem, for
small enough $\epsilon$, the perturbed system~\eqref{eq:main} admits
an asymptotically stable hyperbolic periodic orbit in an
$\epsilon$-neighborhood of $\ket{g}\bra{g}$. The main objective
through the first step of the proof is to characterize this periodic
orbit.

Before going any further and in order to simplify the computations,
we change the language to the Bloch sphere coordinates. Taking
$$
X=\tr{\sigma_x\rho},\quad Y=\tr{\sigma_y\rho},\quad Z=\tr{\sigma_z
\rho},
$$
the system~\eqref{eq:main} reads
\begin{align}\label{eq:blochmain}
\frac{dX}{dt} &= -\Delta
Y-\frac{1}{2}X+\frac{1}{2}(1+Z)X+2\epsilon \kappa_2\cos(\omega t)Z,\\
\frac{dY}{dt} &= \Delta
X-\frac{1}{2}Y+\frac{1}{2}(1+Z)Y-2\epsilon \kappa_1Z,\\
\frac{dZ}{dt} &=
-\frac{1}{2}(1-Z)(1+Z)+2\epsilon\kappa_1 Y-2\epsilon
\kappa_2\cos(\omega t)X.
\end{align}
We proceed the characterization of the periodic orbit through a
perturbative development similar to the Kapitsa shortcut method (see
e.g.~\cite{landau-lifchitz-meca}, page 147). We are looking for a
periodic orbit of the form
$$
 \begin{pmatrix}
    \widetilde X(t) \\
    \widetilde Y(t) \\
    \widetilde Z(t) \\
  \end{pmatrix}
  = \begin{pmatrix}
     0 \\
     0 \\
     -1 \\
    \end{pmatrix}
    + \epsilon  \begin{pmatrix}
    X_1(t) \\
    Y_1(t) \\
    0 \\
  \end{pmatrix}
    + \epsilon^2  \begin{pmatrix}
    X_2(t) \\
    Y_2(t) \\
    Z_2(t) \\
    \end{pmatrix}
    + \epsilon^3  \begin{pmatrix}
    X_3(t) \\
    Y_3(t) \\
    Z_3(t) \\
  \end{pmatrix}
+O(\epsilon^4),
$$
where for the first order approximation, $(X_1,Y_1,Z_1)$, $Z_1$ is
taken to be 0 as the vector must be orthogonal to the unit sphere at
$(0,0,-1)$. Similarly to the Kapitsa method, we choose $X_1(t)$ and
$Y_1(t)$ to be of the form
\begin{equation}\label{eq:kapitsa1}
X_1=\alpha_1\cos(\omega t)+\beta_1\sin(\omega
t)+\gamma_1\quad\text{and}\quad Y_1=\alpha_2\cos(\omega
t)+\beta_2\sin(\omega t)+\gamma_2.
\end{equation}
Inserting~\eqref{eq:kapitsa1} in~\eqref{eq:blochmain}, developing
and considering just the first order terms while regrouping the
$\sin$ and $\cos$ terms, we find the following system:\small
\begin{equation*}
  \left\{
  \begin{aligned}
   \Delta \beta_2 +\frac{1}{2}\beta_1-\omega \alpha_1&=&0\\
  - \Delta \alpha_2 -\frac{1}{2}\alpha_1-\omega
  \beta_1&=&2\kappa_2\\
  - \Delta \beta_1 +\frac{1}{2}\beta_2-\omega \alpha_2&=&0\\
   \Delta \alpha_1 -\frac{1}{2}\alpha_2-\omega \beta_2&=&0
  \end{aligned}
  \right. \qquad\text{and}\qquad   \left\{
  \begin{aligned}
   \Delta \gamma_2 +\frac{1}{2}\gamma_1&=&0\\
  - \Delta \gamma_1 +\frac{1}{2}\gamma_2&=&2\kappa_1
  \end{aligned}
  \right.
\end{equation*}\normalsize
 This system admits for
solution\small
\begin{align}\label{eq:solkap1}
\alpha_1&=-4\kappa_2\frac{4\omega^2+4\Delta^2+1}{\Xi},\quad
\alpha_2=8\kappa_2\Delta\frac{4\omega^2-4\Delta^2-1}{\Xi},\notag\\
\beta_1&=-8\kappa_2\omega\frac{4\omega^2-4\Delta^2+1}{\Xi},\quad
\beta_2=-32\kappa_2\omega\Delta\frac{1}{\Xi},\notag\\
\gamma_1&=-8\kappa_1\Delta\frac{1}{1+4\Delta^2},\quad
\gamma_2=4\kappa_1\frac{1}{1+4\Delta^2},
\end{align}\normalsize
where
$$
\Xi=16\Delta^4+8\Delta^2+1+8\omega^2+16\omega^4-32\Delta^2\omega^2.
$$
Let us go further and consider the second order terms now. Through
the requirement of $\widetilde X^2+\widetilde Y^2+\widetilde Z^2=1$,
one easily has
$$
Z_2(t)=\frac{X_1^2(t)+Y_1^2(t)}{2}.
$$
Moreover, developing the two first equations of~\eqref{eq:blochmain}
up to the second order terms, we have
\begin{align*}
\frac{d}{dt}X_2 &=-\Delta Y_2-\frac{1}{2}X_2,\\
\frac{d}{dt}Y_2 &=\Delta
X_2-\frac{1}{2}Y_2.
\end{align*}
The functions $X_2(t)$ and $Y_2(t)$ being periodic the only
possibility is
$$
X_2(t)=Y_2(t)=0.
$$
Finally, we develop only the third equation of~\eqref{eq:blochmain}
up to the third order terms to obtain
$$
\frac{d}{dt}Z_3=-Z_3,
$$
and as $Z_3$ is periodic the only possibility is $ Z_3(t)\equiv 0$.
Thus we have
$$
\widetilde Z(t)=-1+\epsilon^2\frac{X_1^2+Y_1^2}{2}+O(\epsilon^4).
$$
This yields
\begin{equation}\label{eq:limZ}
\widetilde Z(t)=-1+\epsilon^2 C_1+\epsilon^2 C_2\cos(2\omega
t)+\epsilon^2 C_3\sin(2\omega t)+\epsilon^2 C_4\cos(\omega
t)+\epsilon^2 C_5\sin(\omega t)+O(\epsilon^4),
\end{equation}
where
\begin{align}\label{eq:constants}
C_1&=\frac{\alpha_1^2+\alpha_2^2}{4}+\frac{\beta_1^2+\beta_2^2}{4}+\frac{\gamma_1^2+\gamma_2^2}{2},\notag\\
C_2&=\frac{\alpha_1^2+\alpha_2^2}{4}-\frac{\beta_1^2+\beta_2^2}{4},\qquad
C_3=(\alpha_1\beta_1+\alpha_2\beta_2),\notag\\
C_4&=2(\alpha_1\gamma_1+\alpha_2\gamma_2),\qquad
C_5=2(\beta_1\gamma_1+\beta_2\gamma_2).
\end{align}
This periodic orbit being hyperbolically stable, we have proved the
following lemma:
\begin{lemma}\label{lem:osc}
Consider the system~\eqref{eq:main} with  $\rho(0)=\ket g \bra g$. For any small enough
$\epsilon>0$, there exists a time constant $T>0$ such that
\begin{equation}\label{eq:osc}
\tr{\sigma_z\rho(t)}=\widetilde Z(t) +O(\epsilon^4),\qquad \text{for
} t\geq T,
\end{equation}
where $\widetilde Z$ is given by~\eqref{eq:limZ}.
\end{lemma}

We are now ready to attack the real quantum system with its jumps.
The time constant $T$ in the tuning algorithm of the
subsection~\ref{ssec:algoqubit} is fixed through the
Lemma~\ref{lem:osc}.

\subsection{Step 2: conditional evolution of detuning}

We are interested in  the conditional expectations of $\Delta_{N+1}$
and $\Delta_{N+1}^2$ knowing the value of $\Delta_N$. Due to the
synchronization algorithm $\Delta_{N+1}=\Delta_N-\delta\sin(\omega
t)$, the value of $\Delta_{N+1}$ only depends on the phase
$\phi=\omega t \mod(2\pi)$. We update $\Delta_{N+1}$ only if the
time interval with respect to the previous jump is  large enough to
ensure that the solution of the no-jump dynamics~\eqref{eq:qubit}
has reached its asymptotic regime.  Thus  $(1+Z)/2= \bket{e|\rho|e}
$ is given by~\eqref{eq:limZ} and  the jump probability defined
by~\eqref{eq:jumpqubit} depends only on $\phi=\omega t \mod(2\pi)$.
Since the probability of having a phase $\phi$ during the update
$\Delta_N $ to $\Delta_{N+1}$  is proportional to $\bket{e|\rho|e}
$,  this probability admits a density with respect to the Lebesgue
measure on $[0,2\pi]$, given by
\begin{equation}\label{eq:PphiN}
P_{\varphi,N}= \frac{1}{2\pi}+
\frac{C_2}{2\pi C_1}\cos(2\varphi)+
\frac{C_3}{2\pi C_1}\sin(2\varphi)+
\frac{C_4}{2\pi C_1}\cos(\varphi)+
\frac{C_5}{2\pi C_1}\sin(\varphi)+O(\epsilon^2)
.
\end{equation}
Here the index $N$ denotes the dependence
through~\eqref{eq:constants} and~\eqref{eq:solkap1} of the constants
$\{C_j\}_{j=1,\ldots,5}$ on the detuning $\Delta_N$.

Removing the threshold $C$ in the algorithm by allowing the detuning
to get large, the value of $\Delta_{N+1}$, having fixed $\Delta_N$,
is given as follows
\begin{equation}\label{eq:evolution1}
\Delta_{N+1}=\Delta_N-\delta\sin(\varphi)
\end{equation}
with a probability density
$ P_{\varphi,N}$. Thus
$$
\EE(\Delta_{N+1}~|~\Delta_N) =\Delta_N - \delta
\int_0^{2\pi}\sin(\varphi) P_{\varphi,N}d\varphi =
\Delta_N-\delta\frac{C_5}{2C_1}+O(\epsilon^4) .
$$
Similarly for $\Delta_{N+1}^2$ one has
\begin{equation}\label{eq:evolution2}
\Delta_{N+1}^2=\Delta_N^2-2\delta\sin(\varphi)\Delta_N+\delta^2\sin^2(\varphi)
\end{equation}
with a probability density  $P_{\varphi,N}$.
Inserting~\eqref{eq:PphiN} into~\eqref{eq:evolution2} and with $\delta=\kappa_3 \epsilon^2$,  we have
\begin{align}\label{eq:expectation}
\EE(\Delta_{N+1}^2~|~\Delta_N)&=\Delta_N^2-2\delta\Delta_N\int_0^{2\pi}\sin(\varphi)
P_{\varphi,N}d\varphi+\delta^2 \int_0^{2\pi}\sin^2(\varphi)
P_{\varphi,N}d\varphi
\notag\\
&=\Delta_N^2-\kappa_3\epsilon^{2}\frac{C_5}{C_1}\Delta_N+O(\epsilon^4),
\end{align}
where $\EE(\Delta_{N+1}^2~|~\Delta_N)$ denotes the conditional
expectation of $\Delta_{N+1}^2$ having fixed $\Delta_N$.

Applying~\eqref{eq:constants} and~\eqref{eq:solkap1}, we have
\begin{equation}\label{eq:C5}
C_5=128\kappa_1\kappa_2\frac{\omega(4\omega^2-4\Delta_N^2-1)}{\Xi_N(1+4\Delta_N^2)}\Delta_N.
\end{equation}
with
$$
\Xi_N=16\Delta_N^4+8\Delta_N^2+1+8\omega^2+16\omega^4-32\Delta_N^2\omega^2.
$$
Now, taking into account the threshold $C$ for the growth of the
detuning $\Delta_{N+1}$, applying the assumption~\eqref{eq:eps2:C}
and through some simple computations, we have
\begin{align*}
\frac{4\omega^2-4\Delta_N^2-1}{\Xi_N(1+4\Delta_N^2)} &\geq
\frac{4\omega^2-4C^2-1}{(16\omega^4+8\omega^2+8C^2+1)(1+4C^2)}>0,\\
\alpha_1^2\leq 16\kappa_2^2,~\alpha_2^2\leq 16\kappa_2^2
C^2,&\quad\beta_1^2\leq 64\kappa_2^2\omega^2,~ \beta_2^2\leq
64\kappa_2^2 C^2,\quad \gamma_1^2\leq 16\kappa_1^2,~ \gamma_2^2\leq
16 \kappa_1^2.
\end{align*}
Note in particular that the last line implies
$$
0< C_1\leq 4\kappa_2^2(1+5C^2+4\omega^2)+16\kappa_1^2=:\varrho.
$$
Therefore, we can change the equation~\eqref{eq:expectation} to the
inequality
\begin{equation}\label{eq:ineq}
\EE(\Delta_{N+1}^2~|~\Delta_N)
\leq\Delta_N^2-\epsilon^{2}\varsigma\Delta_N^2+O(\epsilon^4)
\end{equation}
where
\begin{equation}\label{eq:gamma}
\varsigma=64
\kappa_1\kappa_2\kappa_3\omega\varrho\frac{4\omega^2-4C^2-1}{(4C^2+1)\left(16\omega^4+8\omega^2+1+8C^2\right)}>0.
\end{equation}
Taking now the expectation of the both sides of~\eqref{eq:ineq}, we
have
\begin{equation}\label{eq:final}
\EE(\Delta_{N+1}^2)\leq
\left(1-\epsilon^{2}\varsigma\right)\EE(\Delta_N^2)+O(\epsilon^4),
\end{equation}
where we have applied the relation $\EE(\EE(X|Y))=\EE(X)$. Simple computations with
$$
1 + (1-\epsilon^{2}\varsigma) + \ldots  + (1-\epsilon^{2}\varsigma)^N
\leq \frac{1}{\epsilon^2\varsigma}
$$
yield to the  following lemma:
\begin{lemma}\label{lem:expectation}
Considering the Monte-Carlo trajectories described
by~\eqref{eq:qubit}-\eqref{eq:jumpqubit} and applying the
synchronization algorithm of Subsection~\ref{ssec:algoqubit}, we
have
$$
\EE(\Delta_{N}^2)\leq
\left(1-\epsilon^{2}\varsigma\right)^N\Delta_0^2+O(\epsilon^2),
$$
where the positive constant $\varsigma$ is given
in~\eqref{eq:gamma}.
\end{lemma}

This trivially finishes the proof of the Theorem~\ref{thm:qubit} and
we have
$$
\limsup_{N\rightarrow\infty} \EE(\Delta_{N}^2)\leq O(\epsilon^2).
$$
\endproof

\section{Proof of theorem~\ref{thm:lambda}}\label{sec:lambda}

We proceed the proof of the Theorem~\ref{thm:lambda} in a similar
way to that of the Theorem~\ref{thm:qubit}. We assume
$$
1/\omega=\epsilon^2 \kappa_1, \qquad \delta=\epsilon^3\kappa_2,
$$
where $\kappa_1,\kappa_2\sim 1$.

As for theorem~\ref{thm:qubit}, the proof admits  2 main  steps:
\begin{description}
  \item[Step 1] We consider the evolution in the absence of the
  quantum jumps through the system~\eqref{eq:lambda}. We study the
  asymptotic regime of the dynamics. The constant time $T$ will then be chosen to
  ensure the non-jumping system to reach an $\epsilon^3$-neighborhood
  of the limit regime.
  \item[Step 2] In the second step, applying the result of the first step,
  we calculate the conditional expectation of $\Delta_{N+1}$ having fixed $\Delta_N$.
  Finally, we sum up all these results in order to find the limit~\eqref{eq:result3}.
\end{description}

\subsection{Step 1: asymptotic regime of the non-jumping
system}\label{ssec:step1lambda}

We are interested in the dynamics of the system~\eqref{eq:lambda}.
In this aim, we apply the Kapitsa shortcut method. Note that,
\begin{multline*}
\ket{b+\iota\epsilon\cos(\omega t) d}\bra{b+\iota\epsilon\cos(\omega
t) d}
=\ket{b}\bra{b}+\frac{\epsilon^2}{2}\ket{d}\bra{d}\\+\iota\epsilon\cos(\omega
t) (\ket{b}\bra{d}-\ket{d}\bra{b})+\frac{\epsilon^2}{2}\cos(2\omega
t)\ket{d}\bra{d}.
\end{multline*}
Applying the Kapitsa method, the variable $\rho$ may be developed as
\begin{equation}\label{eq:kap}
\rho=\widetilde\rho+O(\frac{\epsilon}{\omega})=\widetilde\rho+O(\epsilon^3),
\end{equation}
where $\widetilde\rho$ represents the unperturbed trajectory. In the
next part, we study the dynamics of the unperturbed part
$\widetilde\rho$.

\subsubsection{Unperturbed no-jump dynamics on the Bloch Sphere}

The unperturbed part, $\widetilde\rho$, satisfies the dynamics:
\begin{multline}\label{eq:unperturbedlambda}
\frac{d}{dt}\widetilde\rho=-\imath\frac{\Delta}{2}\left[\sigma_z,\widetilde\rho\right]
-\frac{1}{2}\left\{\ket{b}\bra{b}+\frac{\epsilon^2}{2}\ket{d}\bra{d},\widetilde\rho\right\}\\
+\tr{\left(\ket{b}\bra{b}+\frac{\epsilon^2}{2}\ket{d}\bra{d}\right)\widetilde\rho}\widetilde\rho.
\end{multline}
In order to study the asymptotic behavior
of~\eqref{eq:unperturbedlambda}, we begin with the case
$\epsilon\equiv 0$ and we study first the system
\begin{equation}\label{eq:unperturbedlambda2}
\frac{d}{dt}\widehat\rho=-\imath\frac{\Delta}{2}\left[\sigma_z,\widehat\rho\right]
-\frac{1}{2}\left\{\ket{b}\bra{b},\widehat\rho\right\}
+\tr{\ket{b}\bra{b}\widehat\rho}\widehat\rho.
\end{equation}
The dynamics in the Bloch sphere coordinates, $X=\tr{\sigma_x
\widehat\rho}$, $Y=\tr{\sigma_y \widehat\rho}$, $Z=\tr{\sigma_z
\widehat\rho}$, are given as follows:
\begin{align*}
\frac{d}{dt} X &= -\Delta
Y-\frac{\sin(2\alpha)}{2}+\left(\frac{\sin(2\alpha)}{2}
X+\frac{\cos(2\alpha)}{2} Z\right) X
\\
\frac{d}{dt} Y &= \Delta X+\left(\frac{\sin(2\alpha)}{2}
X+\frac{\cos(2\alpha)}{2} Z\right) Y
\\
\frac{d}{dt} Z &=-\frac{\cos(2\alpha)}{2}
+\left(\frac{\sin(2\alpha)}{2} X+\frac{\cos(2\alpha)}{2} Z\right) Z
,
\end{align*}
where we have applied $\ket{b}=\cos
\alpha\ket{g_1}+\sin\alpha\ket{g_2}$. Taking
$$
^\prime=2\frac{d}{dt}, \quad p=2\Delta, \quad \beta=2\alpha,
$$
we have the following dynamical system
\begin{equation}
  \label{eq:B}
  \begin{array}{rl}
  X^\prime &= -p Y-\sin\beta+(\sin\beta X+\cos\beta Z) X
\\
Y^\prime &= p X+(\sin\beta X+\cos\beta Z) Y
\\
Z^\prime &=-\cos\beta +(\sin\beta X+\cos\beta Z) Z.
  \end{array}
\end{equation}
living on $\RR^3$.  Since the  two transformations $(X,Y,Z,p,\beta)
\mapsto (-X,-Y,-Z,\beta+\pi)$ and $(X,Y,Z,p,\beta) \mapsto
(X,Y,-Z,\pi-\beta)$ leave the above equations unchanged, we can
always consider, for the study of this dynamical system versus the
parameter $p$ and $\beta$, that the angle
$\beta\in[0,\frac{\pi}{2}]$ and $p\in\RR$.  Since $X^2+Y^2+Z^2=1$ is
invariant, these 3 differential equations define a dynamical system
on the two dimensional sphere $\SSS^2$, the Bloch sphere.

Consider the element of Euclidean length $\delta s^2 = (\delta X)^2
+  (\delta Y)^2  +  (\delta Z)^2$ and its evolution along the
dynamics defined by~\eqref{eq:B} on $\SSS^2$. We have
$$
  \left( \delta s^2\right)^\prime =
2 \left(\delta X \delta X^\prime + \delta Y \delta Y^\prime+  \delta
Z \delta Z^\prime\right)
$$
with $(\delta X^\prime,\delta Y^\prime,\delta Z^\prime)$ given  by
the first variation of~\eqref{eq:B}:
\begin{align*}
 \delta X^\prime &= -p \delta Y +
    (\sin\beta X+\cos\beta Z) \delta X + X (\sin\beta \delta X+\cos\beta \delta Z)
\\
\delta Y^\prime &= p \delta X+
    (\sin\beta X+\cos\beta Z) \delta Y + Y (\sin\beta \delta X+\cos\beta \delta Z)
\\
\delta Z^\prime &=    (\sin\beta X+\cos\beta Z) \delta Z + Z
(\sin\beta \delta X+\cos\beta \delta Z) .
\end{align*}
Since $X \delta X + Y\delta Y + Z \delta Z=0$, we obtain  the simple
relation
\begin{equation} \label{eq:ds2}
     \left( \delta s^2\right)^\prime = 2 (\sin\beta X + \cos\beta Z ) \delta s^2
     .
\end{equation}
Thus $\SSS^2$ splits into two hemispheres: the open hemisphere
$\SSS^2_+$ corresponding to $\sin\beta X + \cos\beta Z >0$ and where
the dynamics is a strict dilation  in any direction; the open
hemisphere  $\SSS^2_-$ corresponding to $\sin\beta X + \cos\beta Z
<0$  where the dynamics is a strict contraction (see~\cite{slotine-auto}).  The boundary
between these two hemispheres is given by the intersection of the
plane $\sin\beta X + \cos\beta Z =0$ with $\SSS^2$. We have
$$
(\sin\beta X + \cos\beta Z)^\prime = -1 - p \sin\beta Y - (\sin\beta
X + \cos\beta Z)^2 .
$$
Thus, when $|p\sin\beta|\leq 1$,  $\SSS^2_+$ is negatively invariant
and $\SSS^2_-$  positively invariant.

Assume first that $p\neq 0$ and $\beta\in ]0,\frac{\pi}{2}[$ and
consider the  equilibrium on $\SSS^2$. Simple computations prove
that  we have only two equilibria associated to the point
$M_+\in\SSS^2_+$ and $M_-\in\SSS^2_-$   of coordinates
$(X_+,Y_+,Z_+)$ and  $(X_-,Y_-,Z_-)$  given by
\begin{align}\label{eq:Mpm}
  \begin{array}{rl}
  X_\pm &= \pm \left(\frac{\cos\beta}{\sin\beta } \right)
    \frac{\sqrt{(p^2-1)^2+4p^2\cos^2\beta} - p^2-1 } {\sqrt{2p^2}  \sqrt{ p^2-1 + \sqrt{(p^2-1)^2+4p^2\cos^2\beta}}}
  \\
  Y_\pm&=\frac{\sqrt{(p^2-1)^2+4p^2\cos^2\beta} -p^2-1 } {2 p\sin\beta}
  \\
  Z_\pm&=\pm \sqrt{ \frac{ p^2-1 + \sqrt{(p^2-1)^2+4p^2\cos^2\beta} } {2p^2} }
  \end{array}
\end{align}
When $p=0$, the above formula  can be extended by continuity to get
the two equilibria:
$$
X_\pm = \pm \sin\beta,\quad   Y_\pm =0, \quad Z_\pm = \pm \cos\beta
.
$$
When $\beta=0$, similarly we obtain the two equilibria
$$
X_\pm = 0,\quad   Y_\pm =0, \quad Z_\pm = \pm 1 .
$$
When $\beta=\frac{\pi}{2}$ the situation is slightly different:
\begin{itemize}
 \item for $|p| < 1$ we have two equilibria
$$
X_\pm =\pm \sqrt{1-p^2} ,\quad   Y_\pm =-p, \quad Z_\pm = 0 .
$$
 \item for $|p| = 1$ we have a unique  equilibrium
$$
X_\pm =0 ,\quad   Y_\pm =-p, \quad Z_\pm = 0 .
$$

 \item for $|p| > 1$ we have two   equilibria
$$
X_\pm =0 ,\quad   Y_\pm =-\frac{1}{p}, \quad Z_\pm = \pm
\sqrt{1-\frac{1}{p^2}} .
$$
\end{itemize}

With all the above  properties we deduce the following lemma
\begin{lemma}\label{lem:bloch}
Consider the differential equations~\eqref{eq:B} defining an
autonomous dynamical system  on the Bloch Sphere  $\SSS^2$ with the
parameters $p\in\RR$ and  $\beta\in[0,\frac{\pi}{2}]$. Then
\begin{enumerate}

\item for $(|p|,\beta)\neq (1,\pi/2)$, we have two distinct equilibrium points $M_+$ and $M_-$ defined here above by~\eqref{eq:Mpm}.
The two Lyapounov exponents at $M_+$ (resp. $M_-$) have strictly
positive (resp. negative) real parts: $M_+$ is locally
exponentially unstable (in all direction) and $M_-$ is locally
exponentially stable.

\item For $|p\sin\beta|< 1$, all the trajectories  (except the unstable equilibrium $M_+$)  converge asymptotically to the  equilibrium point $M_-$
that is exponentially stable: the attraction region of $M_-$ is $\SSS^2/\{M_+\}$.

\end{enumerate}

\end{lemma}

\textit{Proof of Lemma~\ref{lem:bloch}.} The first point result
from~\eqref{eq:ds2} applied locally around $M_+$ and $M_-$ and from
$\sin\beta X_+ + \cos\beta Z_+ >0$ whereas $\sin\beta X_- +
\cos\beta Z_- <0$.

The second point comes from the negative invariance of $\SSS^2_+$,
positive invariance of $\SSS^2_-$ and the Poincare-Bendixon theory
for autonomous systems on the sphere: an hypothetic limit cycle $C$
cannot intersect $\SSS^2_+$ and $\SSS^2_-$ simultaneously and thus
must be included in $\SSS^2_+$ or $\SSS^2_-$; strict surface
dilation (resp. contraction)  in $\SSS^2_+$ (resp. $\SSS^2_-$) is
incompatible with the existence of $C\subset \SSS^2_+$ (resp.
$C\in\SSS^2_-$) because of the Gauss theorem; since there is no
limit cycle and since  there exist only two equilibrium points,
$M_+$ exponentially unstable in all direction and $M_-$
exponentially stable, the attraction domain of $M_-$ is the all
sphere without the unstable point $M_+$.
\endproof

\begin{remark}\label{rem:global}
It is tempting  to conjecture that, for all values of the parameters
$p$ and $\beta$ ensuring two separate equilibria $M_+$ and $M_-$
defined here above, we have a quasi-global convergence towards
$M_-$, the locally exponentially  stable equilibrium.  This is not
true  since for $\beta=\pi/2$ and $|p|>1$ we have the coexistence of
the periodic orbit $X^2+Y^2=1$ with $Z=0$ with the two equilibria
$$
X_\pm =0 ,\quad   Y_\pm =-\frac{1}{p}, \quad Z_\pm = \pm
\sqrt{1-\frac{1}{p^2}}
$$
and thus a trajectory starting with $Z>0$  remains with $Z>0$ for
all the time and cannot converge to $M_-$ since  $Z_- <0$.
\end{remark}

\subsubsection{Perturbed no-jump dynamics}

Under the assumption of the Theorem~\ref{thm:lambda} on $C$, we know
that the detuning $\Delta$ can not get larger than $1/2$ and
therefore in the above notations $p<1$. This trivially implies
$|p\sin\beta|<1$ and therefore we are in the settings of the second
point of the Lemma~\ref{lem:bloch}. Hence, the
system~\eqref{eq:unperturbedlambda2} admits two distinct equilibria
$\overline\rho_-$ and $\overline\rho_+$ given by~\eqref{eq:Mpm} in
the Bloch sphere coordinates. Moreover the trajectories of the
system, not starting at $\overline\rho_+$, necessarily converge
towards the equilibria $\overline\rho_-$.

Applying this characterization of the dynamics, one easily gets
\begin{lemma}\label{lem:asylam2}
Under the assumption of the Theorem~\ref{thm:lambda} for $\ket{b}$
and the  assumption $|\Delta| < \frac{1}{2}$, and for small enough
$\epsilon$, the system~\eqref{eq:unperturbedlambda} admits a locally
asymptotically stable equilibrium $\overline\rho_\epsilon$ of the
form
$$
\overline\rho_\epsilon=\overline\rho_-+\epsilon^2\overline\rho_1+O(\epsilon^4),
$$
where $\overline\rho_-$ is given by~\eqref{eq:Mpm} in the Bloch
sphere coordinates. Moreover the trajectories starting at
$\ket{g_1}\bra{g_1}$ or $\ket{g_2}\bra{g_2}$ converge towards this
equilibrium.
\end{lemma}

For the proof of this lemma, note that, as $\alpha\neq 0$,
$\ket{g_1}\bra{g_1}$ and $\ket{g_2}\bra{g_2}$ are not the
equilibriums of the system~\eqref{eq:unperturbedlambda2}. Thus,
taking $\epsilon$ small enough, they will not be an equilibrium
of~\eqref{eq:unperturbedlambda} neither, and therefore the
trajectories starting at $\ket{g_1}\bra{g_1}$ and
$\ket{g_2}\bra{g_2}$ necessarily converge towards the perturbed
asymptotically stable equilibrium $\overline\rho_\epsilon$.

The Lemma~\ref{lem:asylam2}, together with~\eqref{eq:kap}, implies
that the trajectories $\rho(t)$ of the system~\eqref{eq:lambda}
starting at $\ket{g_1}\bra{g_1}$ or $\ket{g_2}\bra{g_2}$ converge to
an $O(\epsilon^3)$-neighborhood of
$\overline\rho_-+\epsilon^2\overline\rho_1$.

We may therefore choose the time constant $T$ in the synchronization
algorithm of the Subsection~\ref{ssec:algolambda} such that
\begin{equation}\label{eq:asylambda}
\rho(t)=\overline\rho_-+\epsilon^2\overline\rho_1+O(\epsilon^3),\qquad
\forall t>T.
\end{equation}

\subsection{Step 2: conditional evolution of detuning}

Similarly to the last section, we are interested in  the conditional
expectations of $\Delta_{N+1}$ and $\Delta_{N+1}^2$ knowing the
value of $\Delta_N$. Due to the synchronization algorithm
$\Delta_{N+1}=\Delta_N-\delta\sin(2\alpha)\cos(\omega t)$, the value
of $\Delta_{N+1}$ only depends on the phase $\phi=\omega t
\mod(2\pi)$. We update $\Delta_{N+1}$ only if the time interval with
respect to the previous jump is  large enough to ensure that the
solution of the no-jump dynamics~\eqref{eq:lambda} has reached its
asymptotic regime~\eqref{eq:asylambda}.  Thus
$\tr{\ket{b+\imath\epsilon\cos(\omega t)d}
\bra{b+\imath\epsilon\cos(\omega t)d}\rho}$ is given inserting the
limit \eqref{eq:asylambda}. The jump probability defined
by~\eqref{eq:jumptot} depends only on $\phi=\omega t \mod(2\pi)$.
Since the probability of having a phase $\phi$ during the update
$\Delta_N $ to $\Delta_{N+1}$  is proportional to
$\tr{\ket{b+\imath\epsilon\cos(\phi)d}
\bra{b+\imath\epsilon\cos(\phi)d}\rho}$,  this probability admits a
density with respect to the Lebesgue measure on $[0,2\pi]$, given by
\begin{multline}\label{eq:problambda}
P_{\phi,N}=
\frac{1}{\NNN_N(\epsilon)}\Big(\tr{\ket{b}\bra{b}\overline\rho_-}+\epsilon^2\cos^2(\varphi)\tr{\ket{d}\bra{d}\overline\rho_-}
+\epsilon^2\tr{\ket{b}\bra{b}\overline\rho_1}\\
-\epsilon\cos(\varphi)\tr{\sigma_y\overline\rho_-}+O(\epsilon^3)\Big),\qquad\phi\in[0,2\pi),
\end{multline}
where the index $N$ in $P_{\phi,N}$ denotes, in particular, the
dependence of $\overline\rho_-$ and $\overline\rho_1$ to the
detuning $\Delta_N$. Furthermore, the constant $\NNN_N(\epsilon)>0$
is a normalization constant given by the integral over $[0,2\pi]$ of
the term between parentheses. In particular, one easily has
$$
0<O(\epsilon^2)<\NNN_N(\epsilon)<O(1).
$$
Removing the threshold $C$ in the algorithm by allowing the detuning
to get large, the value of $\Delta_{N+1}$, having fixed $\Delta_N$,
is given as follows
\begin{equation}\label{eq:evollam1}
\Delta_{N+1}=\Delta_N-\delta\sin(2\alpha)\cos(\varphi)
\end{equation}
with a probability density $P_{\varphi,N}$.

Similarly for $\Delta_{N+1}^2$ one has
\begin{equation}\label{eq:evollam2}
\Delta_{N+1}^2=\Delta_N^2-2\delta\sin(2\alpha)\cos(\varphi)\Delta_N+\delta^2\sin^2(2\alpha)\cos^2(\varphi),
\end{equation}
with a probability density $P_{\varphi,N}$.

Inserting~\eqref{eq:problambda} into~\eqref{eq:evollam2}, we have
\begin{align}\label{eq:expectationlam}
\EE(\Delta_{N+1}^2~|~\Delta_N)&=\Delta_N^2
-\pi\epsilon\frac{\delta}{\NNN_N(\epsilon)}\frac{\Theta_N}{2}+O\left(\frac{\delta^2}{\NNN_N(\epsilon)}\right)
+O\left(\frac{\delta\epsilon^3}{\NNN_N(\epsilon)}\right),
\end{align}
where
$$
\Theta_N=4\Delta_N^2+1-\sqrt{(4\Delta_N^2-1)^2+16\Delta_N^2\cos^2(2\alpha)}.
$$
Note, in particular, that $\Theta_N>0$ as $\alpha\neq 0$.

Now, taking into account the threshold $C$ for the growth of the
detuning $\Delta_{N+1}$, we can easily see that
$$
\Theta_N=\frac{16\Delta_N^2\sin^2(2\alpha)}{2+16\Delta_N^2\cos^2(2\alpha)+32\Delta_N^4}\geq
\frac{8\sin^2(2\alpha)}{1+8\cos^2(2\alpha)C^2+16C^4}\Delta_N^2.
$$
Therefore, noting by
\begin{equation}\label{eq:varsigma}
\varsigma=\pi\kappa_2\frac{4\sin^2(2\alpha)}{1+8\cos^2(2\alpha)C^2+16C^4}>0,
\end{equation}
where $\delta=\kappa_2\epsilon^3$, we have
\begin{equation}\label{eq:aux2}
\EE(\Delta_{N+1}^2~|~\Delta_N)
\leq\Delta_N^2-~\frac{\epsilon^{4}}{\NNN_N(\epsilon)}\varsigma\Delta_N^2+O(\frac{\epsilon^6}{\NNN_N(\epsilon)}).\\
\end{equation}
Taking now the expectation of the both sides, we have
\begin{equation}\label{eq:finallam}
\EE(\Delta_{N+1}^2)\leq
\left(1-\frac{\epsilon^{4}}{\NNN_N(\epsilon)}\varsigma\right)\EE(\Delta_N^2)+O(\frac{\epsilon^6}{\NNN_N(\epsilon)}),
\end{equation}
where we have applied the relation $\EE(\EE(X|Y))=\EE(X)$. Noting
that
$$
0<O(\epsilon^4)\leq \frac{\epsilon^4}{\NNN_N(\epsilon)}\leq
O(\epsilon^2),
$$
the system~\eqref{eq:finallam} is a contracting one and a similar
computation to that of the last sub-section yields the following lemma:
\begin{lemma}\label{lem:expectationlam}
Considering the Monte-Carlo trajectories described
by~\eqref{eq:lambda}-~\eqref{eq:jump} and applying the
synchronization algorithm of the Subsection~\ref{ssec:algolambda},
we have
$$
\EE(\Delta_{N}^2)\leq
\prod_{k=0}^{N-1}\left(1-\frac{\epsilon^{4}}{\NNN_k(\epsilon)}\varsigma\right)^N\Delta_0^2+O(\epsilon^2),
$$
where the positive constant $\varsigma$ is given
in~\eqref{eq:varsigma}.
\end{lemma}

This trivially finishes the proof of the Theorem~\ref{thm:lambda}
and we have
$$
\limsup_{N\rightarrow\infty} \EE(\Delta_{N}^2)\leq O(\epsilon^2).
$$
Furthermore, note that as the detuning $\Delta_N$ gets near 0, the
normalization constant $\NNN_N(\epsilon)$ converges to an
$O(\epsilon^2)$. This, in particular, leads to a higher convergence
rate in the Lemma~\ref{lem:expectationlam}.
\endproof

\section{Concluding remark}

For 2-level and $\Lambda$-systems, we have proposed synchronization
feedback loops  to lock the probe-frequency to the atomic one.
Simulations illustrate the interest and robustness of such  simple
feedbacks.  Theorems~\ref{thm:qubit} and~\ref{thm:lambda} constitute
a first tentative proving the stability and convergence under
assumptions that seem to be conservative  since they can be relaxed
in simulations. In particular assumptions relative to  100\%
detection efficiently and relative to the minimum  time-delay $T$
between two successive jumps  are not fulfilled  in  simulations of
figures~\ref{fig:qubit} and~\ref{fig:lambda}. We  have observed no
difference when they are satisfied. Thus, we conjecture that
extension of theorems~\ref{thm:qubit} and~\ref{thm:lambda} to
partial detection and $T=0$.

\end{document}